# The Impact of Natural Disasters on Food Security in Türkiye

Raif Cergibozan[a], Emre Akusta[a]

[a] Kırklareli University, Türkiye

**ABSTRACT**

Food security refers to people's access to enough safe nutritious food in order to be able to lead a healthy active life. It also involves elements such as food availability and affordability, as well as people being able to access food that can be consumed healthily. Natural disasters, however, can seriously threaten food security. Disasters' effects on food security are especially more evident in countries such as Türkiye that are frequently exposed to natural disasters due to their geologic and geographical structure. For this reason, the study investigates the effects of natural disasters on food security in Türkiye. The research first creates the Food Security Index in order to estimate the effects of natural disasters on food security. The next phase follows the process of econometric analysis, which consists of three steps. Step one of the econometric analysis uses unit root tests to check the stationarity levels of the series. The second step uses the autoregressive distributed lag (ARDL) bounds test to examine the long-term relationship between natural disasters and food security. The third and final step estimates the effects of natural disasters on food security. According to the obtained results, the study shows earthquakes, storms, and floods to have a significant short- as well as long-term negative effect on food security. The overall impact of natural disasters on food security has also been determined to be negative.

Food security is a basic concept that refers to people's access to enough safe nutritious food in order to be able to lead a healthy active life. This concept has vital importance in terms of meeting basic needs and has preserved its importance for humanity throughout history. Food security has changed and developed based on factors such as agricultural productivity, resistance to natural disasters, and the effectiveness of food distribution systems. These days, however, food security has taken on a complex dynamic structure under the impact of factors such as the increasing global population, climate change, economic uncertainties, and natural disasters. Ensuring food security is closely related not just to protecting individuals' right to adequate nutrition but also to sustainably increasing social welfare.

First brought to the agenda at a conference held in the US state of Virginia in 1943, food security began to take form through factors such as food supply shortages and people's inability to obtain sufficient sources of nutrition. The concept of food security has evolved over time into a complex structure that comprises not just food availability but also a number of components such as access to food, economic status, social inequalities, and environmental factors (Shetty, 2015).

The Food and Agriculture Organization (FAO, 2018, p. 3) of the United Nations defines food security as "a situation that exists when all people, at all times, have physical, social and economic access to sufficient, safe and nutritious food that meets their dietary needs and food preferences for an active and healthy life." This definition emphasizes food security to include such dimensions as not only food availability but also accessibility, usability, and stability. These dimensions must be fulfilled simultaneously in order to ensure food security. These four dimensions should also be safeguarded over time in order to ensure long-run food security. Meanwhile, among the other factors affecting food security are resilience to shocks and disasters, social safety nets, and sustainable agricultural practices (Dragica & Nikolić, 2020).

**CORRESPONDENCE TO:** Raif Cergibozan (PhD), Department of Economics, Kırklareli University, Kırklareli 39010 Türkiye.
Email: raif.cergibozan@klu.edu.tr ORCID: 0000-0001-7557-5309

Food security refers to food accessibility and affordability, as well as people's ability to consume foods healthily. When looked at from this perspective, the effects of natural disasters on food security can seriously threaten these basic elements. For example, damage to agricultural land, reduced harvests, and food shortages in the event of a flood or drought can negatively impact food access and costs. Similarly, the infrastructural damage and logistical problems in the aftermath of an earthquake can hinder food distribution and limit people's access to food. Understanding natural disasters' impact on food security will provide the opportunity to take effective measures against these risks and to respond faster in crisis situations. Strengthening disaster management policies and agricultural production systems in accordance with these risks carries great importance, especially in countries such as Türkiye that are frequently exposed to natural disasters. For this reason, the study examines the effects of natural disasters on food security in Türkiye. In line with this, the study will evaluate the effects disasters have on food production, distribution, and access and will present recommendation regarding the necessary measures to be taken.

When looking at studies on natural disasters, the vast majority of studies in the literature are seen to have examined the effects of natural disasters on the economy and economic indicators (For examples, see Cavallo & Noy, 2009; Schumacher & Strobl, 2011; Arouri et al., 2015; Liu et al., 2020; Qin et al., 2021). Studies on the economic effects of natural disasters have generally examined their effects on such economic indicators as external debt, growth, budget deficit, and inflation. Among these studies, Murlidharan and Shah (2001) used regression analysis to investigate the relationship between disasters and the growth of external debt in 32 developed and developing countries between 1980-1995. Kahn's (2005) study concluded that the government, a strong financial sector, higher education, and social institutions are important for reducing the impact of natural disasters. Raddatz (2007) examined natural disasters' effect on short-run output volatility in low-income countries and found natural disasters to have a negative impact. Rodriguez-Oreggia (2010) analyzed the effects of natural events on the Human Development Index. According to the obtained results, natural disasters cause poverty to increase and the Human Development Index to decrease. Meanwhile, Noy and Nualsri (2007) used a panel vector autoregression (VAR) analysis to estimate the economic consequences of natural disasters. They found natural events to have a negative impact on government budgets, with government spending increasing and taxes decreasing following a major disaster event.

Meanwhile, studies investigating the effects of natural disasters on economic growth have indicated major disasters to have negative impacts on short- and long-run output (e.g., Cavallo et al., 2013). While Skidmore and Toya (2002) argued educated countries with high income to experience fewer losses from natural disasters, Loayza et al. (2012) stated that disasters affect economic growth but that the effects vary according to the type of disaster and economic sectors. While some moderate disasters can have positive growth impacts on certain sectors, severe disasters generally do not. Moreover, growth in developing countries is more sensitive due to the sensitive nature of their economic sectors. Another area of research involves the effects of natural disasters on agriculture. Again, Loayza et al. (2012) found disasters have negative effects on agriculture, with droughts and storms in particular negatively impacting agriculture. Sivakumar (2005) stated that natural disasters damage agriculture and that this causes hunger and poverty. Long (1978) argued natural disasters to have caused an increase in hunger and poverty in many low-income countries.

Research on how to prevent and reduce the economic impacts of disasters has focused on strategies such as updating warning systems, increasing preparedness for natural disasters, and reducing disaster losses. Toya and Skidmore (2007) and Padli et al. (2018) determined sustainable economic development to be able to decrease disaster risks, while Noy (2009) argued better institutions, higher literacy, and higher trade to be able to regulate the effects of disasters. Taghizadeh-Hesary et al. (2019) showed infrastructure improvements to be able to reduce disaster losses, while Padli and Habibullah (2009) found an increase in human development to reduce disaster losses. Alongside these studies, Abbas et al. (2018) studied flood prevention strategies in Pakistan. Their study developed a flood disaster index involving social, economic, and psychological impacts that showed improving housing levels, disaster preparedness, and being away from rivers to reduce flood damage. They also emphasized the importance of adopting resilience strategies such as early warning systems, preflood planning, compensation policies, and information about regional flood intensity. Studies conducted to identify the determinants of food security (e.g., Noy, 2009; Padli et al., 2018) found factors such as farm size, income level, and female education level to affect food security.

In conclusion, although the literature has numerous studies on the economic impacts of natural disasters, the relationship between food security and natural disasters has yet to be sufficiently researched. In addition, no empirical study is found to have investigated the effects of natural disasters on food security in Türkiye. Therefore, this study investigates the effects of natural disasters on food security in Türkiye, thus aiming to fill this gap in the literature. Firstly, the Food Security Index was created for this purpose. The next phase follows the process of econometric analysis, which consists of three steps. The first step of the econometric analysis uses unit root tests to check the stationarity levels of the series. The second step uses the autoregressive distributed lag (ARDL) bounds test to examine the long-run relationship between natural disasters and food security. The third and final step estimates the impact of natural disasters on food security within the framework of four separate models.

## The Relationship Between Natural Disasters and Food Security

Natural disasters have complex and wide-ranging impacts on agriculture, food security, and natural resources. In order to evaluate these impacts, the direct and indirect effects generally need to be examined separately. Figure 1 provides a detailed framework for analyzing the impacts natural disasters have on agriculture, food security, and natural resources from this perspective.

When assessing natural disasters' effects on agriculture, the direct effects are separated into positive and negative effects. Positive direct effects include increased water supply, increased soil fertility, and the expansion of aquatic habitats. Meanwhile, the negative direct effects include factors such as reduced output; damage to agricultural inputs, facilities, and/or infrastructure; limited planting options; damage to farm supply routes and markets; and death or injury to agricultural workers. Additionally, the indirect effects include factors such as increased production costs, lower agricultural productivity, decreased food supply, and higher food prices (Israel & Briones, 2012).

The direct and indirect effects are discussed in detail while evaluating the effects of natural disasters on natural resources and the environment. Positive direct effects include factors such as higher humidity and less air pollution. Negative direct effects, however, include soil erosion, drought, silting and sedimentation, reduced tree and vegetation cover, reduced soil fertility, an accumulation of waste and water pollution, salt water intrusion, high tides, storm waves, and deformation of land topography. Indirect effects include factors such as the reduced resilience of ecosystems and the endangerment of human health and safety (Israel & Briones, 2012).

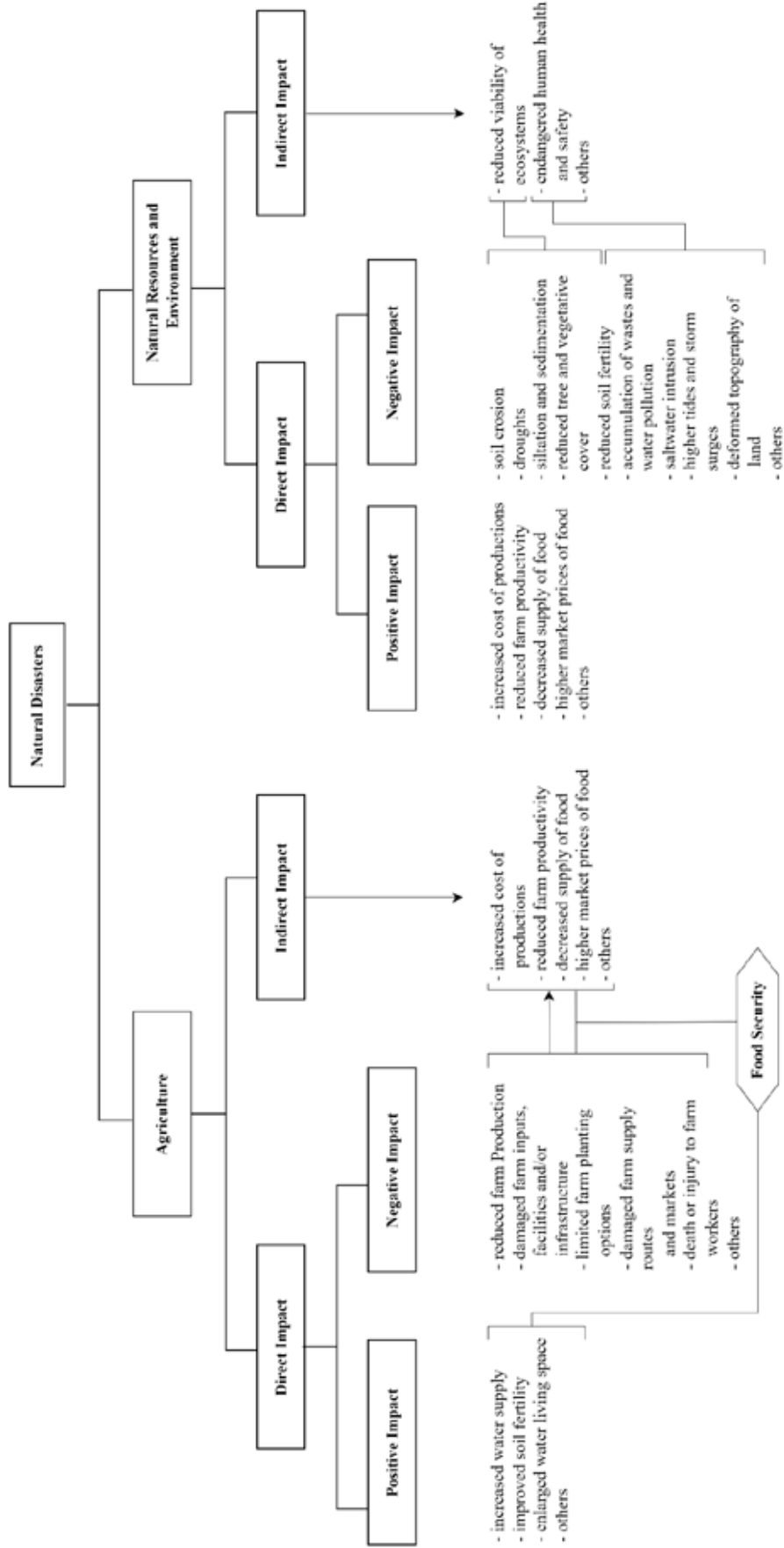

*Figure 1.* Analysis framework regarding the impacts of natural disasters on agriculture, food security, natural resources, and the environment (Israel & Briones, 2012, p. 4).

When evaluating the effects on the agricultural, natural resource, and the environmental sectors together, positive and negative effects are seen to occur hand in hand, but the net effect is assumed to be negative in general. However, one should not forget that in some cases, the negative as well as positive effects of natural disasters can be observed. Therefore, a holistic approach is required to understand and manage the complex impacts natural disasters have on agriculture, food security, and natural resources.

**Natural Disasters and Their Effects in Türkiye**

Türkiye is a country that is frequently exposed to natural disasters due to its geological and geographical structure. Over 95% of the country lies in one of the world's most active earthquake and landslide zones. When considering the impact natural disasters have had in Turkey over the last 70 years, more than 100,000 people are seen to have lost their lives, more than 600,000 buildings to have been damaged, and more than 500,000 buildings to have been affected by earthquakes in various ways. These data reveal the country to be at risk of disaster and this risk to be significant. Earthquakes especially are the type of natural disaster Türkiye experiences most frequently that have the most devastating consequences (Istanbul Technical University [ITU], 2023). In this context, the effects of natural disasters, in particular floods, droughts, and earthquakes, have great importance on food security in Türkiye. The frequency and severity of natural disasters can directly affect food security. Disasters can reduce food supplies by damaging agricultural land and production capacity, which in turn can lead to increased food prices, difficulties with accessibility, and malnutrition. Therefore, understanding the effects of natural disasters and developing effective strategies for coping with them is vital in order for Türkiye to maintain food security. For this purpose, Figure 2 examines the numerical distribution of natural disasters in Türkiye by year.

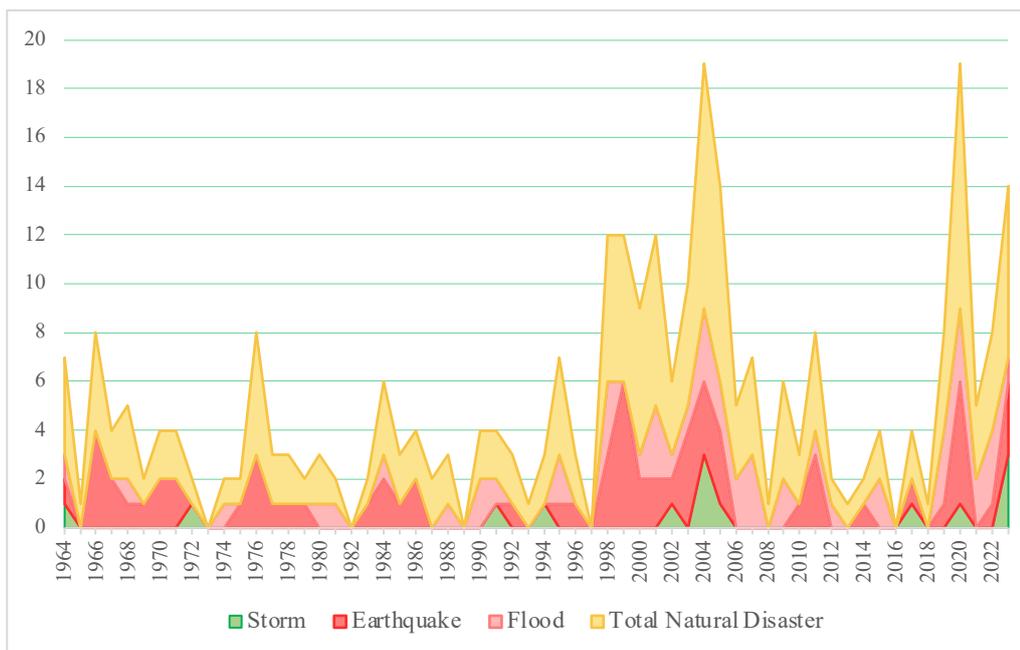

*Figure 2.* Distribution of natural disasters in Türkiye over time (Formed from data from the Centre for Research on the Epidemiology of Disasters' [CRED] International Disasters Database [EM-DAT]).

When examining Figure 2, while only a few natural disasters are seen to be experienced in some years, up to 19 occur in other years. In the late 1990s and early 2000s, a significant increase occurred in the number of natural disasters. During this period, 1999 was particularly important as this was when Türkiye faced events such as earthquakes and floods. In recent years, in particular 2020, 2022, and 2023, natural disasters occurred very frequently. Meanwhile, examining the effects and damage natural disasters cause is very important in addition to examining them quantitatively. For this reason, Figure 3 shows the number of people who've lost their lives or been affected by the natural disasters that have occurred in Türkiye in recent times, as well as the economic damage these disasters have caused. Figure 3 has total deaths read on the primary (left) axis, while total number of those affected and total damage are read from the secondary (right) axis.

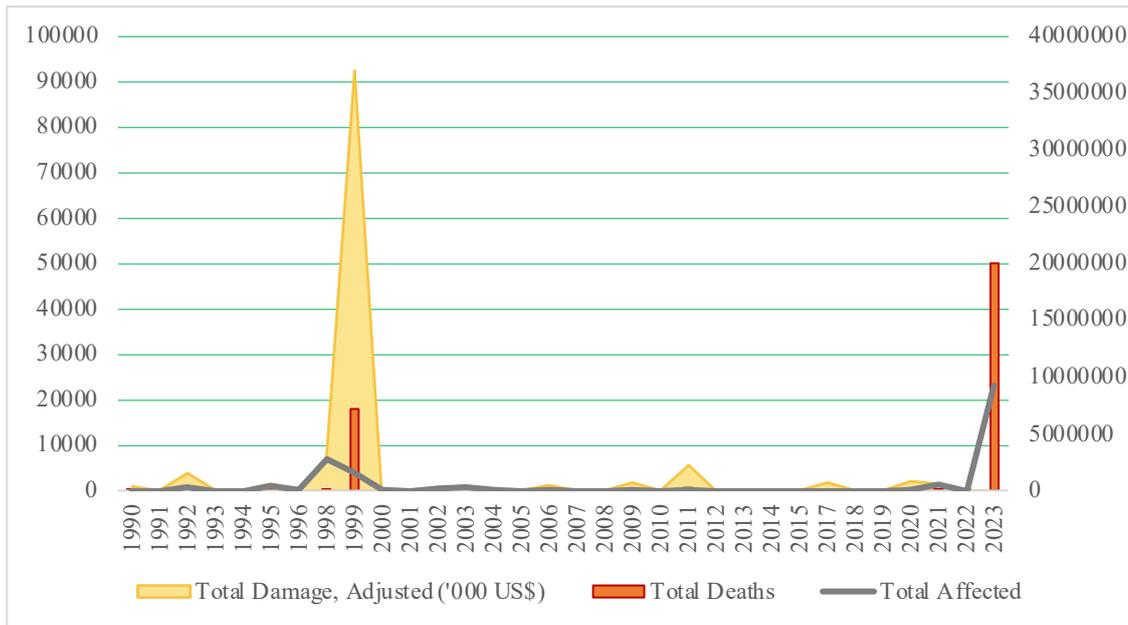

*Figure 3.* The economic and social impacts of natural disasters (formed from EM-DAT data).

When examining the data in Figure 3, the 1999 and 2023 earthquakes are particularly striking. Large-scale earthquakes such as the 1999 Marmara Earthquake and the 2023 Kahramanmaraş-centered earthquakes show how sensitive Türkiye is to such disasters. These disasters have also pushed the country's capacity for coping with natural disasters to its limits. According to the EM-DAT data, a total of 17,982 people lost their lives in the Marmara Earthquake, while this number grew to 50,135 for the Kahramanmaraş-centered earthquake. When looking at the total number of people affected, while 1,584,792 people were affected by the Marmara Earthquake, 9,208,252 are registered as having been affected by the Kahramanmaraş-centered earthquake. Additionally, while the economic damage from the Marmara Earthquake was reported as around $37 billion, the economic damage from the Kahramanmaraş-centered earthquake has yet to be officially announced. However, some studies have estimated it to be $70 billion, others $103 billion, and others still $150 billion (Çubukcu et al., 2023; ITU, 2023; Özer, 2023).

These data also show that Türkiye is at risk from earthquakes and that it must continue to improve its ability to cope with this type of disaster. After the 1999 earthquake, a trend in decreasing earthquake-related losses was observed in Türkiye alongside an increase in preventative measures and structural regulations. However, large-scale earthquakes such as the 2023 earthquakes highlight that the country is still vulnerable to natural disasters and that more improvements and preparations are needed. As a result, more comprehensive and effective strategies need to be developed and implemented for coping with future disasters. Increasing the levels of earthquake awareness and preparedness and strengthening preventative measures and emergency response plans are particularly important. This is critical both for protecting human life as well as minimizing physical damage.

When examining the effects of natural disasters, earthquakes in particular, the challenges Türkiye faces become clear. However, these disasters affect more than just human life; they also deeply impact important areas such as agriculture, food security, and natural resources. The agricultural sector is one that has felt feels these effects most directly and clearly. Disasters such as earthquakes, floods, and storms can cause serious damage to agricultural fields and negatively impact crop output. Alongside this, the damage to agricultural infrastructure can also cause long-run effects. When looked at from the perspective of natural resources and the environment, disasters are seen to cause problems such as soil erosion, drought, and water pollution. Floods can especially cause the erosion of agricultural lands and soil fertility to decrease, while factors such as water pollution can negatively impact natural life and ecosystems. When considering the effects of natural disasters on agriculture, food security, and natural resources, these disasters clearly pose a serious issue. Therefore, developing and implementing disaster management and risk reduction policies have great importance. Resilience in the agricultural and environmental sectors also needs to be increased, and disaster preparedness needs to happen. Alongside reducing the effects of natural disasters, these steps will also help Türkiye build a more sustainable future.

**Data and Methodology**

The study first constructed the Food Security Index in order to estimate the effects natural disasters have on food security. The next stage follows a three-step econometric analytical process. The first step of the econometric analysis uses unit root tests to check the stationarity levels of the series. The second step uses the ARDL bounds test to examine the long-run relationship between natural disasters and food security. The third and final step estimates the impact of natural disasters on food security within the framework of four separate models. The details of the methods used are given below.

**Model Specifications and Data**

This research is an empirical study that investigates the effects of natural disasters on food security in Türkiye. The study uses annual data from the 1990-2022 period. The research model can be expressed functionally as in Equation 1.

$$FSI_{(M-D)} = f(GDP, TRADE, GENGOV, CURRACC, GROSSFIX, AGRCLTR, FOREST, TOTAL) \tag{1}$$

In Equation 1, *FSI* is the dependent variable of the research model and represents the Food Security Index. Khan et al.'s (2022) approach was followed when creating the Food Security Index. Normalization of the data was achieved using the min-max method, and the index was created using principal component analysis (PCA). Table 1 presents the indicators used in the Food Security Index.

Table 1

*Indicators of the Food Security Index*

| Indicators | Unit | Data source |
| --- | --- | --- |
| Access to electricity, rural | % of rural population | WB-WDI (2024) |
| Agriculture, forestry, and fishing, value added | % of GDP | WB-WDI (2024) |
| Grain yield | Kilograms per hectare | WB-WDI (2024) |
| Employment in agriculture | % of total employment | WB-WDI (2024) |
| Food production index | 2014–2016 =100 | WB-WDI (2024) |
| Livestock production index | 2014–2016 =100 | WB-WDI (2024) |
| Rural population | % of total population | WB-WDI (2024) |
| Rural population growth | Annual % | WB-WDI (2024) |

WB-WDI = World Bank-World Development Indicators

Data on natural disasters were obtained from the EM-DAT database maintained by CRED. The natural disasters on which our study reports include earthquakes, storms, floods, and other natural disasters. The study follows Noy's (2009) approach for measuring the impacts of natural disasters. The natural disaster metrics are calculated using three damage metrics: (1) the number of people affected (injured, homeless, or otherwise affected), (2) the number of fatalities, and (3) the amount of economic damage (direct amount of damage in USD). These measurements are expressed below (Benali et al., 2019, p. 699), with the calculations shown respectively in Equations 2, 3, and 4.

$$\text{Total population affected} = \left[\sum_{t=1}^{T} \left(\frac{\text{total population affected}_{ijt}}{\text{total population}_{i,t}}\right)\right] \tag{2}$$

$$\text{Total fatalities} = \left[\sum_{t=1}^{T} \left(\frac{\text{total people killed}_{ijt}}{\text{total population}_{i,t}}\right)\right] \tag{3}$$

$$\text{Economic Damage} = \left[\sum_{t=1}^{T} \left(\frac{\text{Damage}_{ijt}}{\text{total GDP}_{i,t}}\right)\right] \tag{4}$$

The variable on the right side of Equation 1 are the determinants of food security as identified based on previous literature (see Israel & Briones, 2012; Sekhampu, 2013; Khan et al., 2022; Ma et al., 2023). These eight variables are common control variables for all the models. The main explanatory variable in the research model is total damage from natural disasters (*TOTAL*). Total damage from natural disasters is addressed within the framework of the four separate models in order to estimate the impact of each natural disaster separately. The first three models have been established to estimate the impacts of the most common natural disasters in Türkiye (i.e., earthquakes, storms, floods) on food security. The last model, Model 4, shows the total impact of natural disasters. These four models take place in Equations 5-8.

$$\text{Model (1): } FSI_{(M-D)} = \alpha_0 + \beta_1 GDP_t + \beta_2 TRADE_t + \beta_3 GENGOV_t + \beta_4 CURRACC_t + \beta_5 GROSSFIX_t + \beta_6 AGRCLTR_t + \beta_7 FOREST_t + \beta_8 QUAKE_t + \mu_t \quad (5)$$

$$\text{Model (2): } FSI_{(M-D)} = \alpha_0 + \beta_1 GDP_t + \beta_2 TRADE_t + \beta_3 GENGOV_t + \beta_4 CURRACC_t + \beta_5 GROSSFIX_t + \beta_6 AGRCLTR_t + \beta_7 FOREST_t + \beta_8 STORM_t + \mu_t \quad (6)$$

$$\text{Model (3): } FSI_{(M-D)} = \alpha_0 + \beta_1 GDP_t + \beta_2 TRADE_t + \beta_3 GENGOV_t + \beta_4 CURRACC_t + \beta_5 GROSSFIX_t + \beta_6 AGRCLTR_t + \beta_7 FOREST_t + \beta_8 FLOOD_t + \mu_t \quad (7)$$

$$\text{Model (4): } FSI_{(M-D)} = \alpha_0 + \beta_1 GDP_t + \beta_2 TRADE_t + \beta_3 GENGOV_t + \beta_4 CURRACC_t + \beta_5 GROSSFIX_t + \beta_6 AGRCLTR_t + \beta_7 FOREST_t + \beta_8 TOTAL_t + \mu_t \quad (8)$$

In the econometric models, is the constant terms and expresses the cut-off point. ,, , ..., are the slope coefficients for the variable they appear alongside. is the error term of the models, which are assumed to have a mean of zero and constant variance, while the *t* sub-index represents the time period. The research uses annual data from the 1990-2022 period, and the data for the variables are secondary data that have been obtained from different databases. Table 2 shows how the variables are defined, the databases from which they were taken, and some descriptive statistics such as means, minima, maxima, and standard deviations.

Table 2

*Descriptive Statistics of the Research Variables (1990-2022)*

| Variable | Notation | *M* | *Mdn* | Min. | Max. | *SD* | Source |
| --- | --- | --- | --- | --- | --- | --- | --- |
| Food Security Index | *FSI* | 6.06E-09 | -0.56 | -1.22 | 1.67 | 1.00 | Calculated from WB data. |
| Damage From Earthquakes | *QUAKE* | 0.01 | -0.38 | -0.38 | 4.35 | 1.14 | Calculated from WB data. |
| Damage From Storms | *STORM* | -0.12 | -0.14 | -0.14 | 0.35 | 0.09 | Calculated from WB data. |
| Damage From Floods | *FLOOD* | 0.15 | -0.18 | -0.18 | 7.53 | 1.37 | Calculated from WB data. |
| Damage From Natural Disasters | *TOTAL* | -0.02 | -0.26 | -0.28 | 6.98 | 1.26 | Calculated from WB data. |
| Forest area (% of land area) | *FOREST* | 27.07 | 26.91 | 25.71 | 29.07 | 1.08 | WB-WDI (2024) |
| GDP per capita (constant 2015 US$) | *GDP* | 8382 | 7863 | 5257 | 14055 | 2666 | WB-WDI (2024) |
| Trade (% of GDP) | *TRADE* | 49.20 | 48.76 | 30.48 | 81.17 | 10.88 | WB-WDI (2024) |
| General government final consumption expenditure (% of GDP) | *GENGOV* | 13.08 | 12.91 | 10.61 | 15.66 | 1.32 | WB-WDI (2024) |
| Current account balance (% of GDP) | *CURRACC* | -2.72 | -3.07 | -8.87 | 2.01 | 2.57 | WB-WDI (2024) |
| Gross fixed capital formation (% of GDP) | *GROSSFIX* | 25.49 | 25.98 | 17.95 | 29.86 | 3.21 | WB-WDI (2024) |
| Agricultural land (% of land area) | *AGRCLTR* | 51.14 | 51.15 | 49.01 | 53.56 | 1.43 | WB-WDI (2024) |

**Index Techniques**

Multivariate statistical analysis aims to explain *p* number of variables belonging to *n* number of objects. However, the large number of variables, as well as the relationships between variables, make qualifying and understanding the analyses difficult. PCA is used in such cases as it allows a large number of variables to be explained with fewer independent variables. PCA is the most established and well-known technique in multivariate statistical analysis (Jolliffe, 2002). This technique was first introduced by Pearson (1901), developed by Hotelling (1933), and has been widely used by many disciplines over time.

PCA is used to obtain a smaller number of variables that are describable as a combination of variables from a large number of variables that interact with one another. In this way, no statistical data losses occur, and subjective weighting is avoided. Also, *p* number of variables, where *n* is the number of measurements, are transformed into *k* number ($k \square p$) of new variables that are linear and independent from one another (Joliffe, 2002, p. 167).

The first step when starting PCA is to normalize the variable data. This is because the different value ranges for each attribute cause distortions of attributes with much smaller values than the other attributes. Methods such as min-max, z-score, decimal scaling, sigmoid, and softmax normalizations are used for transforming data (Organisation for Economic Co-operation & Development, 2008). This study uses the min-max normalization method for normalizing the data.

**Unit Root Analysis**

The variables used in time series analyses must be stationary in order to obtain reliable results for forecasting. However, economic time series often have non-stationary characteristics (Johansen & Jesulius, 1990). Having non-stationary variables in time series analyses causes the problem of spurious regression, and estimates become biased and inconsistent (Granger & Newbold, 1974). Meanwhile, the level at which variables become cointegrated determines which cointegration estimation method to use. For this reason, the first step in a time series analysis is to use unit root tests to check the stationarity of each variable.

This study uses the ARDL method. While this method permits explanatory variables to be cointegrated (stationary) at different levels, the dependent variable must be stationary at the I(0) level and any of the explanatory variables to be non-stationary at the second-level difference, or I(2). For this purpose, the first stage of the analysis determines the stationarity levels of the series using the augmented Dickey-Fuller (ADF) and Phillips-Peron (PP) unit root tests. The basic hypothesis for both tests is that the series has a unit root. Therefore, the underlying hypothesis is rejected when the calculated test statistics exceed the critical values.

**ARDL Bounds Test**

The second stage of the analysis uses the ARDL bounds test proposed by Pesaran et al. (2001) to determine the presence of a cointegration relationship between variables. This method has been frequently preferred in similar studies for its significant advantages over traditional cointegration estimation methods. Some of the advantages of the ARDL bounds test are as follows: Firstly, unlike traditional cointegration tests, the stationarity levels such as I(0) or I(1) of the series can differ. Secondly, the lag lengths of the variables included in the model can also differ. Thirdly, multiple parameter estimates can be made with a single ARDL equation. Fourthly, short- and long-run coefficients can be estimated with the help of reduced equations. Fifthly, ARDL enables consistent estimates to be made, even in cases with small samples and endogeneity problems. Due to these features, the ARDL model has an important place in the literature (Pesaran et al., 2001; Narayan & Smith, 2005; Topal & Şentürk, 2019). A two-variable ARDL(*p*, *q*) model is expressed in Equation 5 (Cergibozan et al., 2017):

$$(5) \quad y_t = a_0 + \sum_{i=1}^{p} \varphi_i y_{t-i} + \beta' x_t + \sum_{i=0}^{q-1} \beta^{*'} \Delta x_{t-i} + u_t$$

where the vector represents the different *k*-dimensional variables that are not cointegrated with one another; represents the error term with a mean of zero, constant variance, and no autocorrelation. When rearranging the model with polynomial delay operators and , the simplified form of the equation is shown in Equation 6.

$$(6) \quad C(L)y_t = \mu + B(L)X_t + \delta W_t + \varepsilon_t$$

ARDL bounds testing is performed by estimating the error correction model in the ARDL model using the least squares (LSS) method. Equation 7 shows the cointegration test equation based on the ARDL($p$, $q$) technique.

$$(7) \quad \Delta y_t = \alpha_0 + \pi_1 y_{t-1} + \pi_2 x_{t-1} + \sum_{i=1}^{p-1} \psi_i^2 \Delta y_{t-i} + \sum_{i=0}^{q} \zeta_i \Delta x_{t-i} + \varepsilon_t$$

where $\Delta$ represents the difference operator. is the constant term, and $p$ and $q$ represent the lag lengths of the dependent and independent variables, respectively. The F-test is used to determine whether a cointegration relationship is present or not. The F-test statistic is calculated as shown in Equation 8.

$$(8) \quad F_{BDM} = \frac{(SSR_r - SSR_{ur})/r}{SSR_{ur}/(T-k)}$$

where SSRr and SSRur are the sum of the squared errors obtained from the estimation of the restricted and unrestricted models, respectively; r and (T - k) are the degrees of freedom, with r representing the number of restrictions, $T$ is the total number of observations, and $k$ represents the number of parameters estimated in the unrestricted model (Cergibozan et al., 2017, p. 75). The basic hypothesis of the test is that the coefficients from the first lag of the raw values of the variables are equal to zero (; in other words, no cointegration relationship exists between the variables. Deciding whether the basic hypothesis can be rejected occurs by using the F-test statistic to compare the critical values Pesaran et al. (2001) tabulated for the lower bound I(0) and upper bound I(1) sets, respectively. In order to reject the main hypothesis, the F-statistic must be greater than the upper bound I(1) values at a statistical significance level of at least 10%. If the test statistic is less than the lower bound I(0) levels, the basic hypothesis cannot be rejected. If the test statistic falls between the lower and upper limits, different cointegration tests should be used to check for the presence of a cointegration relationship.

**ARDL Long-Term Estimation**

Because the cointegration relationship was found significant as a result of the F-test, the last stage of the analysis makes long-run coefficient estimates. Equation 9 shows the long-term equation that will be estimated in the two-variable ARDL($p$, $q$) model.

$$(9) \quad \Delta y_t = \alpha_0 + \sum_{i=1}^{p} \varphi_i \Delta y_{t-i} + \sum_{i=0}^{q} \zeta_i \Delta x_{t-i} + \varepsilon_t$$

ARDL estimates are very sensitive to lag length. Pesaran et al. (2001) stated that the Akaike (AIC) or Schwarz (SBC) information criteria should be used to determine the optimum lag length. This study uses AIC to determine the appropriate lag lengths. In order for the ARDL estimates to be considered accurate, the model's error term must also be normally distributed, autocorrelated, and have constant variance; the functional form must be constructed properly, the estimated parameters must be stable, and diagnostic test must be used to test the basic assumptions. Short- and long-run coefficient estimates can be taken into account when the F-statistic is significant and the assumptions are valid.

**Empirical Findings**

The first stage of the analysis applied the ADF and PP unit root tests to the series of variables in the research models in order to determine their stationarity levels. Tables 3 and 4 show the estimation results of the unit root models including the intercept term and the intercept and trend terms, respectively.

Table 3

*Unit Root Test Results (Intercept)*

| Variables | ADF Unit Root Tests | | PP Unit Root Tests | |
|---|---|---|---|---|
| | *t*-statistic (level) | *t*-statistic (first difference) | *t*-statistic (level) | *t*-statistic (first difference) |
| *FSI* | -1.303 | -4.934*** | -1.303 | -4.941*** |
| *QUAKE* | -2.420 | -3.140*** | -3.366** | -10.567*** |
| *STORM* | -5.773*** | -9.327*** | -5.778*** | -30.433*** |
| *FLOOD* | -1.526 | -9.187*** | -5.822*** | -15.607*** |
| *TOTAL* | -5.640*** | -8.719*** | -5.641*** | -29.977*** |
| *FOREST* | -0.096 | -1.212*** | 2.651 | -2.748*** |
| *GDP* | 1.908 | -4.721*** | 4.213 | -4.669*** |
| *TRADE* | -0.046 | -4.770*** | 0.415 | -4.576*** |
| *GENGOV* | -2.397 | -5.979*** | -2.424 | -6.044*** |
| *CURRACC* | -3.580** | -9.045*** | -3.644*** | -13.145*** |
| *GROSSFIX* | -1.864 | -5.854*** | -1.864 | -5.922*** |
| *AGRCLTR* | -0.677 | -4.545*** | -0.875 | -4.545*** |

*Note.* ***, **, and * indicate a 1%, 5%, and 10% level of significance, respectively.

Table 4

*Unit Root Test Results (Trend and Intercept)*

| Variables | ADF Unit Root Tests | | PP Unit Root Tests | |
|---|---|---|---|---|
| | *t*-statistic (level) | *t*-statistic (first difference) | *t*-statistic (level) | *t*-statistic (first difference) |
| *FSI* | -1.062 | -5.119*** | -1.159 | -5.153*** |
| *QUAKE* | -3.139 | -8.286*** | -3.388* | -10.515*** |
| *STORM* | -5.677*** | -9.171*** | -5.679*** | -31.612*** |
| *FLOOD* | -3.618* | -17.077*** | -6.022*** | -15.450*** |
| *TOTAL* | -5.706*** | -15.103*** | -5.754*** | -30.405*** |
| *FOREST* | -3.205 | -0.903*** | -2.460 | -1.537*** |
| *GDP* | -1.1692 | -5.318*** | -0.418 | -7.998*** |
| *TRADE* | 0.499 | -4.731*** | -1.662 | -4.595*** |
| *GENGOV* | -2.276 | -6.003*** | -2.276 | -6.060*** |
| *CURRACC* | -4.203** | -8.891*** | -4.310*** | -14.216*** |
| *GROSSFIX* | -2.617 | -5.761*** | -2.686 | -5.821*** |
| *AGRCLTR* | -1.761 | -4.463*** | -1.920 | -4.473*** |

*Note.* ***, **, and * indicate a 1%, 5%, and 10% level of significance, respectively.

When examining the results from the ADF unit root tests, the basic hypothesis claiming a unit root for STORM, TOTAL, and CURRACC variables is seen to be rejected in both the intercept, trend, and intercept models, albeit at different statistical significance levels, while the basic hypothesis cannot be rejected for the series regarding the other variables. The PP unit root test results resemble the ADF results, with the only difference being that the QUAKE variable does not contain a unit root. Meanwhile, according to both unit root test results, the first differences for all series were determined to be stationary. Lastly, the dependent variable of the research models, the $FSI_{(M-D)}$ series, was determined to be non-stationary at the level regarding both tests and models, but became stationary when taking the first difference, $I(1)$. According to the results from the unit root tests, examining the cointegration relationship with the ARDL bounds test instead of traditional cointegration tests was determined to be appropriate because the research models' dependent variable was $I(1)$ and the explanatory variables were $I(1)$ or $I(0)$.

The second stage of the analysis applies the ARDL bounds test to determine whether the long-run relationship between the variables is significant. Table 5 shows the results from the ARDL bounds test that was applied to the four different research models.

Table 5

*ARDL Bounds Test Results*

| Model | Optimum Delay Length | F-statistic | Critical Value 5% | | Critical Value 1% | |
|---|---|---|---|---|---|---|
| | | | $I(0)$ | $I(I)$ | $I(0)$ | $I(I)$ |
| Model 1: F(*FSI* | *QUAKE, FOREST, GDP, GENGOV, GROSSFIX, AGRCLTR*) | (1, 0, 0, 0, 0, 0, 0, 0) | 7.390*** | 2.27 | 3.28 | 2.88 | 3.99 |
| Model 2: F(*FSI* | *STORM, FOREST, GDP, GENGOV, GROSSFIX, AGRCLTR*) | (1, 0, 0, 0, 0, 0, 1, 0, 0) | 8.116*** | 2.11 | 3.15 | 2.62 | 3.77 |
| Model 3: F(*FSI* | *FLOOD, FOREST, GDP, GENGOV, GROSSFIX, AGRCLTR*) | (1, 0, 0, 0, 0, 0, 0, 0, 0) | 8.222*** | 2.11 | 3.15 | 2.62 | 3.77 |
| Model 4: F(*FSI* | *TOTAL, FOREST, GDP, GENGOV, GROSSFIX, AGRCLTR*) | (1, 0, 1, 0, 1, 1, 1, 0, 0) | 10.541*** | 2.11 | 3.15 | 2.62 | 3.77 |

*Note.* *** and ** indicate 1% and 5% significance levels, respectively.

The first column of Table 5 contains the function forms of the cointegration models, the second column contains the optimal lag lengths determined using AIC, the third column contains the F-statistics calculated to test the basic hypothesis that no cointegration is present, and the last columns contain the critical values at the statistical significance levels of 5% and 1%. According to the results, due to the F-test statistics calculated for all four models being greater than the limit values, the basic hypothesis claiming no cointegration is rejected. In other words, a statistically significant long-run relationship was found among the variables in all four models.

Due to the ARDL bounds test results providing evidence of a long-run relationship between the explanatory variables and the dependent variable ($FSI_{(M-D)}$), the third and final stage of the analysis estimates the long-run coefficients. Table 6 shows the long-run estimation results. The bottom of the table shows the results from the diagnostic tests determining whether the estimated parameters show stability. According to the results from the diagnostic tests, the error terms of all three models are normally distributed, have no autocorrelations, and have constant variance; the functional form of the models are shown to have been correctly constructed and the estimated parameters to be stable. According to the results from the diagnostic tests, the estimated F-statistics and slope coefficients for the models are consistent. In addition, Figure 4 presents the graphs for the cumulative sum (CUSUM) and cumulative sum of squares (CUSUMSQ) statistics created for each model.

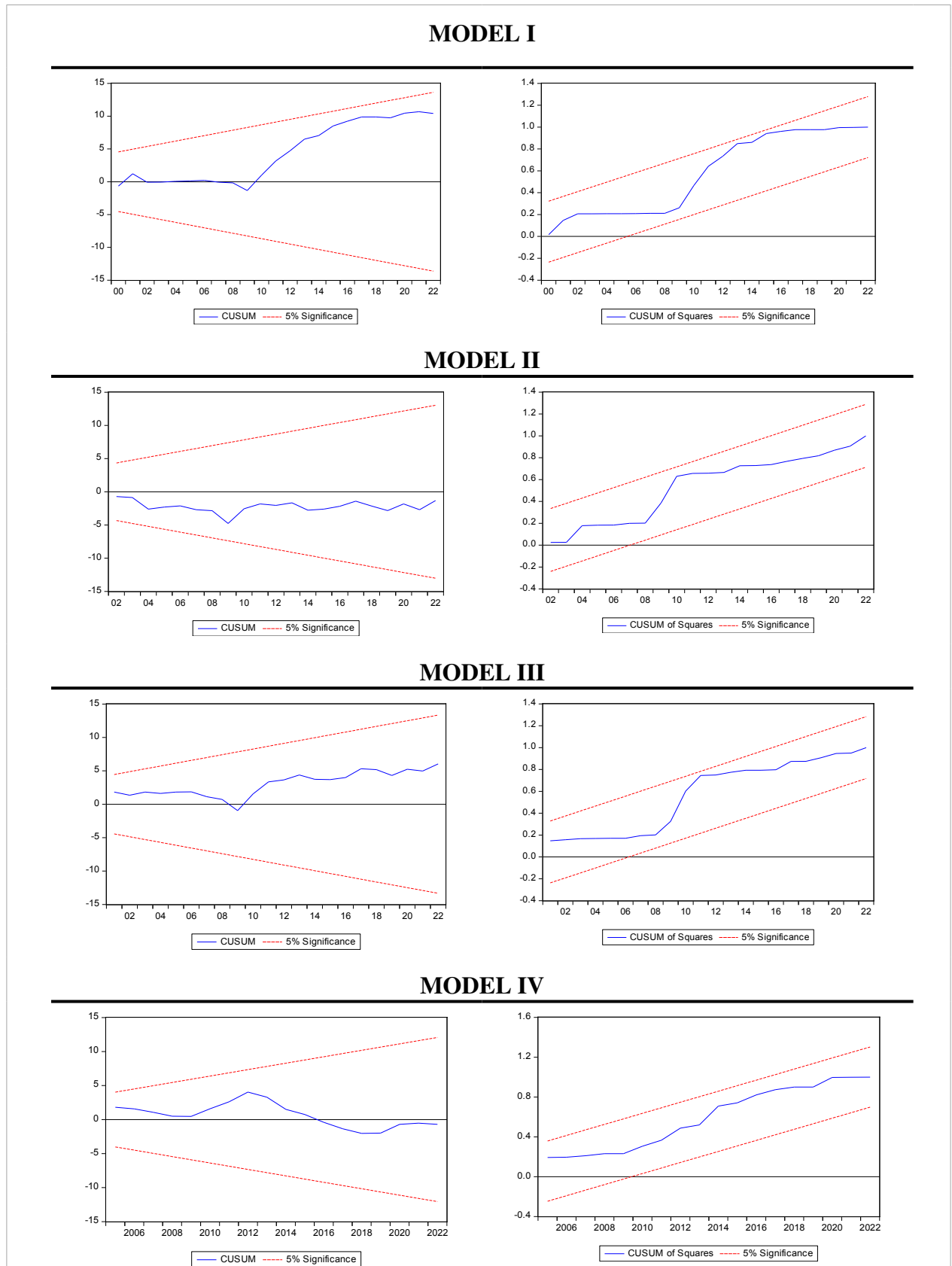

*Figure 4.* Plots for the CUSUMSQ and CUSUM statistics.

Table 6

*Short- and Long-Run Analysis Results*

| Dependent Variable: $FSI_{(M-D)}$ | Short-Run Coefficients | | | |
|---|---|---|---|---|
| Explanatory Variables | Model 1 | Model 2 | Model 3 | Model 4 |
| QUAKE | -0.035** | | | |
| STORM | | -0.277** | | |
| FLOOD | | | -0.026** | |
| TOTAL | | | | -0.015* |
| FOREST | -0.362*** | 0.210 | 0.082 | -0.323** |
| GDP | -0.001** | -0.001** | -0.001** | -0.001** |
| TRADE | -0.007** | 0.003 | -0.002 | -0.002*** |
| GENGOV | -0.154*** | -0.100*** | -0.136*** | -0.133*** |
| CURRACC | 0.030*** | 0.022*** | 0.026*** | 0.030*** |
| GROSSFIX | -0.014* | -0.008 | -0.012 | -0.011 |
| AGRCLTR | -0.184*** | -0.113*** | -0.152*** | -0.176*** |
| ECT(-1) | -0.483*** | -0.301*** | -0.377*** | -0.488*** |
| Dependent Variable: $FSI_{(M-D)}$ | Long-Run Coefficients | | | |
| Explanatory Variables | Model 1 | Model 2 | Model 3 | Model 4 |
| QUAKE | -0.057* | | | |
| STORM | | -0.107** | | |
| FLOOD | | | -0.097*** | |
| TOTAL | | | | -0.132*** |
| FOREST | -0.793*** | 0.467 | -0.074 | -0.694** |
| GDP | 0.001 | -0.001 | -0.001 | 0.001 |
| TRADE | -0.030*** | -0.028** | -0.018* | -0.031*** |
| GENGOV | -0.317*** | -0.365*** | -0.384*** | -0.306*** |
| CURRACC | 0.060*** | 0.085*** | 0.081*** | 0.059*** |
| GROSSFIX | -0.118*** | -0.042 | -0.051* | -0.112*** |
| AGRCLTR | -0.408*** | -0.449*** | -0.471*** | -0.407*** |
| C | 50.041*** | 21.096*** | 35.571*** | 47.366*** |
| Diagnostic tests | *p* value | *p* value | *p* value | *p* value |
| $\chi^2$ *(Serial correlation)* | 0.17 | 0.20 | 0.14 | 0.47 |
| $\chi^2$ *(Heteroscedasticity)* | 0.71 | 0.64 | 0.42 | 0.74 |
| $\chi^2$ *(Normality)* | 0.76 | 0.65 | 0.88 | 0.69 |
| $\chi^2$ *(Functional form)* | 0.43 | 0.18 | 0.24 | 0.32 |
| CUSUM | Stable | Stable | Stable | Stable |
| CUSUMSQ | Stable | Stable | Stable | Stable |

*Note.* *** and ** indicate a 1% and 5% level of significance respectively.

According to the estimation results in Table 6, the error correction term for all models was found to be negative and statistically significant. These results show that a long-run relationship exists between the variables and that if an imbalance occurs, this balance will be restored. The error correction term in Table 6 indicate that approximately 48% of the deviations occurring between the short and long-run in Model 1, approximately 30% in Model 2, approximately 38% in Model 3 and approximately 49% in Model 4 will be eliminated in the next period. The short- and long-run coefficients are generally seen to be significant.

Models 1, 2, and 3 were respectively established to estimate the impacts of earthquakes, storms, and floods on food security. When looking at the results from these three models, earthquakes, storms, and floods are seen to have significant negative short- and long-run effects on food security. In addition, the other factors used in the analysis appear to have varying effects on food security. For example, forest areas have a negative impact on food security, meaning that as forest areas increase, food security decreases. Factors such as trade volume as a percentage of GDP and general government expenditures as a percentage of GDP also have negative effects, meaning that as these factors increase, food security decreases. Meanwhile, current account balance is seen to have a positive impact on food security.

Model 4 was established to measure the total impact of natural disasters on food security. According to the long-run results from Model 4, natural disasters negatively impact food security. The loss of life and property caused by natural disasters also causes a decrease in the Food Security Index. In addition, natural disasters can disrupt food supply chains, damage storage and transportation systems, and damage local agricultural outputs. In this situation, food supplies may decrease or their quality may deteriorate. Another problem that can appear is clean water sources becoming polluted. These are the reasons why natural disasters negatively affect food security. Forest areas increasing also negatively affects food security. These obtained results resemble those from the studies of Long (1978), Sivakumar (2005), and Loayza et al. (2012). As forest areas increase, agricultural lands may decrease or restrictions can be imposed on agricultural activities. Decreasing agricultural land can lead to a decrease in food production capacity, thus posing a risk to food security. In addition, restrictions or difficulties regarding agricultural activities can also lead to restrictions in terms of efficiency and diversity in agricultural production, which may reduce food supply and negatively affect food security.

While no significant effect on food security was detected regarding an increase in GDP levels, food security was found to decrease as trade increased. An increase in trade is thought to be able to negatively impact local food production and reduce food security. Several factors must be considered in order to understand how an increase in trade affects local food production. Firstly, an increase in trade often means an increase in imports. Competition from imported food products can reduce local producers' share of the market and lead to a decrease in local production. In this case, food security risks may come about unless additional measures are taken to increase local producers' competitiveness. Meanwhile, how an increase in trade affects the balance of supply and demand regarding food production is also important. An increase in trade generally occurs due to an increase in the demand for foreign trade. This may cause local producers to turn to exports and to reduce their output in order to meet the demand from local markets. As a result, local food supplies may decrease and food security may become jeopardized. The impact an increase in trade has on food prices should also be taken into account. Price fluctuations in foreign trade and an increase in imports may affect local food prices and cause food costs to increase. This could jeopardize food security, especially for low-income groups. However, an increase in the current account balance as a percentage of GDP leads to an increase in food security. The current account surplus means that a country receives more foreign exchange. This foreign exchange can be used to import food. The current account surplus can also strengthen a country's economy, by increasing food production and distribution, helping the country pay off its debts, and preventing debt crises that threaten food security. These results resemble those in the studies by Noy and Nualsri (2008) and Padli and Habibullah (2009).

## Results

Food security is a fundamental concept that refers to people having access to sufficient, safe, and nutritious food in order to lead a healthy and active life. While this concept has vital importance for meeting the basic needs of humanity, it is not just limited to protecting individuals' right to adequate nutrition. It is also closely related to sustainably increasing social welfare. Meanwhile, natural disasters are an important factor affecting food security. The effects of disasters on food security are more evident, especially in countries such as Türkiye that are frequently exposed to natural disasters due to their geological and geographical structure.

The frequency and severity of natural disasters in Turkey is able to directly affect food security. Natural disasters such as earthquakes, floods, storms, and droughts are able to cause serious damage to agricultural lands and negatively affect crop output. These disasters are also able to cause damage to agricultural infrastructure, which leads to long-run impacts. The agricultural sector is one that feels the effects of natural disasters most directly and clearly. Damage to agricultural lands can lead to a decrease in food production capacity, which can pose a risk to food security. For this reason, the current study has investigated the effects of natural disasters on food security in Türkiye. In order to estimate the effects of natural disasters on food security, the research started by creating the Food Security Index. The next stage used a time series analysis to estimate the effects of natural disasters on food security. Meanwhile, this study has some limitations. The study can be improved by resolving these limitations in future studies. Firstly, this study has used annual data from the 1990-2022 period. Secondly, the study used four models and eight explanatory variables in each model. The depth of the study can be increased by expanding the study's period and dataset. Thirdly, the study sample has been limited to Türkiye. The sample of subsequent studies can be expanded, or the results can be compared by selecting developed and developing countries.

Models 1, 2, and 3 in the study were set up to respectively estimate the effects of earthquakes, storms, and floods on food security. As the last model, Model 4 shows the total impact of natural disasters. According to the obtained results, the study has shown earthquakes, storms, and floods to have significant negative impacts on food security, both in the short- and long-run. Natural disasters as a whole have also been found to have a negative impact on food security.

Countries such as Türkiye that are frequently exposed to natural disasters need to take strategic and effective steps in order to minimize their impact on food security. The results and findings from this study's analyses highlight the significant negative impacts natural disasters have on food security. Therefore, various measures need to be taken in order to be prepared for future natural disasters and to maintain food security: *(1) Strengthen Disaster Management:* Strong disaster management policies and planning are vital for an effective fight against natural disasters. Reviewing and updating existing disaster management plans can ensure rapid and effective intervention before, during, and after a disaster. In addition, mechanisms should be established to encourage cooperation among local governments, emergency teams, and civil society organizations. In this way, rapid coordinated actions can be taken in disaster situations. *(2) Strengthen Early Warning Systems:* Improved early warning systems are critical for identifying disaster risks in advance and informing the public in a timely manner. The technological infrastructure of these systems should be strengthened and enabled to reach every segment of society. Raising public awareness and training against disasters can increase the effectiveness of early warning systems. Regular training programs should also be organized on how to act in disaster situations. *(3) Strengthen the Agricultural Substructure:* Equipping agricultural lands with water and erosion control systems can reduce the effects disasters have on agricultural output. Infrastructure projects such as dams, reservoirs, and irrigation systems can ensure the sustainability of agricultural production. In addition, providing post-disaster support to farmers is important so that agricultural production can be restarted quickly and economic losses can be minimized. These types of support may include various mechanisms such as financial and technical assistance, credit assistance, and agricultural insurance. *(4) Promote Alternative Agricultural Methods:* Alternative farming methods such as organic farming, greenhouse farming, and hydroponic farming can reduce the effects natural disasters have on agricultural production. Encouraging and supporting these methods is an important step toward maintain food security. In addition, greater investments should be made in agricultural practices that combat climate change and promote environmental sustainability. *(5) Disaster Insurance and Financial Measures:* Financial measures such as disaster insurance should be provided to farmers, and necessary incentives should be given to implement these. Disaster insurance can help farmers compensate for the losses natural disasters cause and ensure continuity of agricultural output. In addition, state-supported disaster funds should be established to provide rapid and effective financial assistance to disaster-affected regions. These funds can accelerate post-disaster recovery and minimize economic losses. These policy recommendations could be important steps toward increasing Türkiye's resilience against natural disasters and ensuring food security. In order for these measures to be implemented effectively, however, cooperation and coordination should be ensured among the public, private sector, civil society organizations, and international stakeholders.

**Authors' contribution**

Author Contributions: Conception/Design of study: R.C.; Data Acquisition: R.C., E.A.; Data Analysis/Interpretation: R.C.; Drafting Manuscript: R.C., E.A.; Critical Revision of Manuscript: R.C., E.A.; Final Approval and Accountability: R.C., E.A.

**Peer-review**

Externally peer-reviewed

**Funding**

This project was supported by the Kırklareli University, Scientific Research Fund BAP Project No. KLÜBAP-240.

**Disclosure statement**

The authors report no conflict of interest.

**Author's ORCID numbers**

| Raif Cergibozan | 0000-0001-7557-5309 |
| --- | --- |
| Emre Akusta | 0000-0002-6147-5443 |